# Universal momentum-to-real-space mapping of topological singularities

Xiuying Liu[1+], Shiqi Xia[1+], Ema Jajtić[2+], Daohong Song[1,3*], Denghui Li[1], Liqin Tang[1,3], Daniel Leykam[4], Jingjun Xu[1,3], Hrvoje Buljan[1,2*], and Zhigang Chen[1,3,5*]

1The MOE Key Laboratory of Weak-Light Nonlinear Photonics, TEDA Applied Physics Institute and School of Physics, Nankai University, Tianjin 300457, China
2Department of Physics, Faculty of Science, University of Zagreb, Bijenička c. 32, 10000 Zagreb, Croatia
3Collaborative Innovation Center of Extreme Optics, Shanxi University, Taiyuan, Shanxi 030006, People's Republic of China
4Center for Theoretical Physics of Complex Systems, Institute for Basic Science, Daejeon 34126, Republic of Korea
5Department of Physics and Astronomy, San Francisco State University, San Francisco, California 94132, USA
*Corresponding authors: songdaohong@nankai.edu.cn, hbuljan@phy.hr, zgchen@nankai.edu.cn
+These authors made equal contribution.

**Topological properties of materials, as manifested in the intriguing phenomena of quantum Hall effect and topological insulators [1,2], have attracted overwhelming transdisciplinary interest in recent years[3-7]. Topological edge states, for instance, have been realized in versatile systems including electromagnetic-waves[8-12]. Typically, topological properties are revealed in momentum space, using concepts such as Chern number and Berry phase. Here, we demonstrate a universal mapping of the topology of Dirac-like cones from momentum space to real space. We evince the mapping by exciting the cones in photonic honeycomb (pseudospin-1/2) [13,14] and Lieb (pseudospin-1) lattices[15] with vortex beams of topological charge $l$, optimally aligned for a chosen pseudospin state $s$, leading to direct observation of topological charge conversion that follows the rule of $l \to l + 2s$ (see Figs. 1a, 1b). The mapping is theoretically accounted for all initial excitation conditions with the pseudospin-orbit interaction and nontrivial Berry phases. Surprisingly, such a mapping exists even in a deformed lattice where the total angular momentum is not conserved, unveiling its topological origin. The universality of the mapping extends beyond the photonic platform and 2D lattices: equivalent topological conversion occurs for 3D Dirac-Weyl synthetic magnetic monopoles[16-18] (see Fig. 1c), which could be realized in ultracold atomic gases[19] and responsible for mechanism behind the vortex creation in electron beams traversing a magnetic monopole field[20].**

The coupling of spin and orbital degrees of freedom is in many systems intertwined with the underlying topology of the space and the Berry phase[21]. For instance, in condensed matter electronic systems, study of spin-orbit interaction leads to discovery of topological insulators, which have emerged as an important field for itself. The physics of electron beams illustrates many examples where spin-orbit coupling is integrated with topology[22]. There is also a plethora of related examples in optics[23]: with real space Berry phase optical elements such as q-plates and metasurfaces, circular polarization of light (intrinsic spin) can be transformed to an optical vortex carrying orbital angular momentum (OAM) [24-26]; for light propagating along a coiled ray trajectory, the dynamics is governed by the action of the monopole in Berry curvature, leading to the spin-Hall effect of light[27]. Interestingly, an analogous topological transport of sound waves was recently observed, thanks to the spin-redirection geometric phase[28].

When discussing spin in optical systems, it is light polarization or photon spin that is usually considered as the spin degree of freedom[23, 29]. Similarly, in electronic systems it is the intrinsic electron spin[1,2]. However, for light (electrons) propagating in structured photonic media (crystalline lattices) with inherent degrees of freedom, the concept of pseudospin independent of any intrinsic particle property emerges[13-15, 30-32]. Since angular momentum is defined abstractly via the set of commutation relations for corresponding operators, pseudospin should be treated on equivalent footing as other angular momenta in a given system. Consequently, a whole class of fundamental phenomena based on pseudospin-orbit interaction, twined together with topology of the underlying space, should be expected in photonic, electronic, and other relevant platforms with emergent pseudospins. We demonstrate here one such phenomenon: momentum-to-real-space mapping of topological singularities. This is achieved, experimentally, by proper excitation of pseudospin states near Dirac-like cones in 2D photonic honeycomb and Lieb lattices (as illustrated in Figs. 1a and 1b). We develop a unified theory to explain the observed phenomenon, extending the mapping to the 3D Dirac-Weyl systems where synthetic monopoles in momentum space (as illustrated in Fig. 1c) play an essential role.

The honeycomb lattice (HCL)[33] is composed of two triangular sublattices (*A*, *B*), while the Lieb lattice[34-36] has three square sublattices (*A*, *B, C*). The HCL features conical intersections with two touching bands at two inequivalent Dirac points (*K* and *K'*), representing a pseudospin-1/2 system ($S = 1/2$, Fig. 1a). In contradistinction, the Lieb lattice possesses a conical intersection with three touching bands at the Dirac-like *M* points, representing a pseudospin-1 system ($S = 1$, Fig. 1b). For excitations around conical intersections in both lattices, the dynamics is governed by the Hamiltonian

$$H = \kappa(S_x k_x + S_y k_y), \qquad (1)$$

where $S_i$ are the components of the pseudospin angular momentum operator **S**, and $\kappa$ depends on the properties of the lattice. The eigenstates of the pseudospin $\chi_{S,s}$ are given by $\mathbf{S}^2 \chi_{S,s} = S(S+1)\chi_{S,s}$, and $S_z \chi_{S,s} = s\chi_{S,s}$ (here *S* and *s* denote the total and *z*-component pseudospin angular momentum, respectively). The eigenmodes of the Hamiltonian, $H\psi_{n,\mathbf{k}} = \beta_{n,\mathbf{k}}\psi_{n,\mathbf{k}}$, are organized in $2S + 1$ bands (labeled by *n*) touching at the conical intersection. In contrast to previous excitation schemes, we use multiple (three for the honeycomb and four for the Lieb) vortex beams (each with an initial topological charge $l = 1$ or $l = -1$) momentum-matched to the conical intersection points (see Figs. 1a, 1b), spatially structured to excite only one pseudospin eigenstate ($s = -1/2$ or $1/2$ for the HCL, and $s = -1, 0$ or $1$ for the Lieb lattice). Further experimental details about lattice creation in a biased nonlinear SBN crystal and excitation scheme can be found in the Supplementary Material (SM).

Typical experimental results obtained with the HCL are summarized in Fig. 2. The HCL (Fig. 2a) is established with the multi-beam optical induction technique[13]. It remains invariant throughout a 20-mm-long crystal with a nearest neighbor spacing of 9 μm. The lattice is probed by a donut-shaped triangular lattice beam, for which the OAM ($l$) and pseudospin ($s$) are optimally aligned (top panel: $l = 1, s = 1/2$, bottom panel $l = -1, s = -1/2$). To see better the phase structure of the probe beam at input (i.e., before the pseudospin-orbit interaction takes place), interferograms are obtained for the whole superimposed beam (Fig. 2b) as well as one of the three interfering beams (Fig. 2c). As illustrated in Fig. 1a, the vortex beams are momentum-matched to the three equivalent Dirac points ($K$) of the HCL. The output interferograms in Figs. 2d and 2e clearly display two vortices of the same helicity, which show conversion of the topological charges from $l$ to $l + 2s$ for both initial states (schematically illustrated in Fig. 1a). We demonstrate below that this conversion is a consequence of the mapping of topological singularity at the conical intersections from the momentum to the real space. Figure 2e is obtained from the Fourier transform of spectral component at one of the Dirac points back into real space for phase measurement. The bottom-right inset in Fig. 2e shows a donut-shaped intensity pattern at the output, which is somewhat deformed compared to that at the input (see the inset in Fig. 2c) due to instability of high-order vortices and inhomogeneity of the nonlinear crystal.

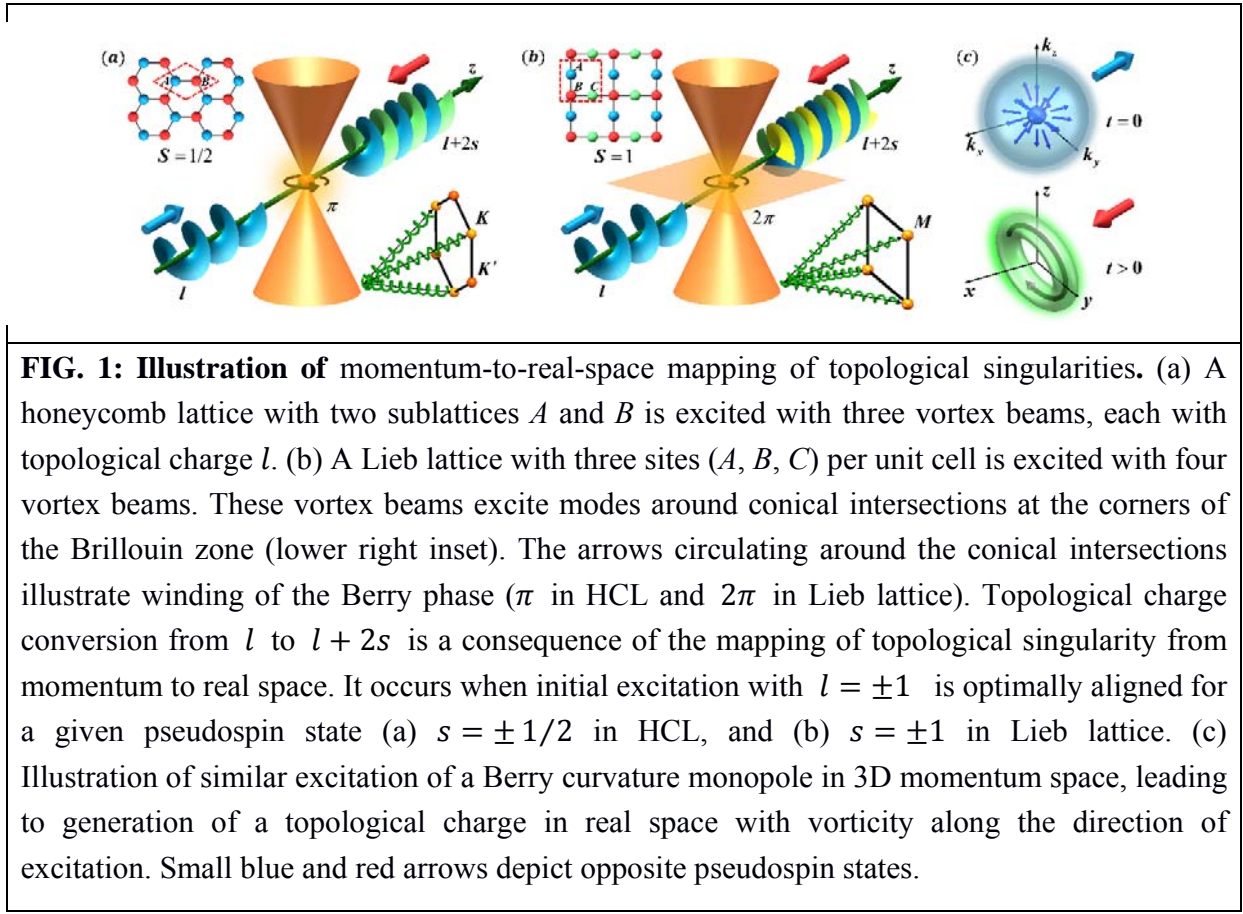

**FIG. 1: Illustration of** momentum-to-real-space mapping of topological singularities. (a) A honeycomb lattice with two sublattices $A$ and $B$ is excited with three vortex beams, each with topological charge $l$. (b) A Lieb lattice with three sites ($A$, $B$, $C$) per unit cell is excited with four vortex beams. These vortex beams excite modes around conical intersections at the corners of the Brillouin zone (lower right inset). The arrows circulating around the conical intersections illustrate winding of the Berry phase ($\pi$ in HCL and $2\pi$ in Lieb lattice). Topological charge conversion from $l$ to $l + 2s$ is a consequence of the mapping of topological singularity from momentum to real space. It occurs when initial excitation with $l = \pm 1$ is optimally aligned for a given pseudospin state (a) $s = \pm 1/2$ in HCL, and (b) $s = \pm 1$ in Lieb lattice. (c) Illustration of similar excitation of a Berry curvature monopole in 3D momentum space, leading to generation of a topological charge in real space with vorticity along the direction of excitation. Small blue and red arrows depict opposite pseudospin states.

It is instructive to provide a kinematical explanation of these observations (see SM). In the experiment of Fig. 2, the $z$-component $J_z$ of the total angular momentum $\mathbf{J} = \mathbf{L} + \mathbf{S}$ is conserved (where $\mathbf{L} = \mathbf{r} \times \mathbf{k}$ is the OAM): $[J_z, H] = 0$. The initial excitation in all experiments is comprised of a single value of $l$ and $s$, and the optimally aligned initial condition implies maximal value of $|l + s|$. The output beam has two or more values of $l'$ and $s'$, all of which obey

$$l + s = l' + s'. \qquad (2)$$

For example, for Fig. 2 – top panel, the input beam has $l = 1$ and $s = 1/2$, while the output beam has two components: (i) $l' = 1$ and $s' = 1/2$, and (ii) $l' = 2$ and $s' = -1/2$. Since the output components are intertwined on both sublattices, we observe two vortices in Figs. 2d and 2e. Fully equivalent explanation holds for results in Fig. 2 – bottom panel. However, as we shall discuss below through strict theoretical anaylsis, the observed charge conversion has a topological origin, holding even in systems without rotational symmetry in which angular momentum is not conserved.

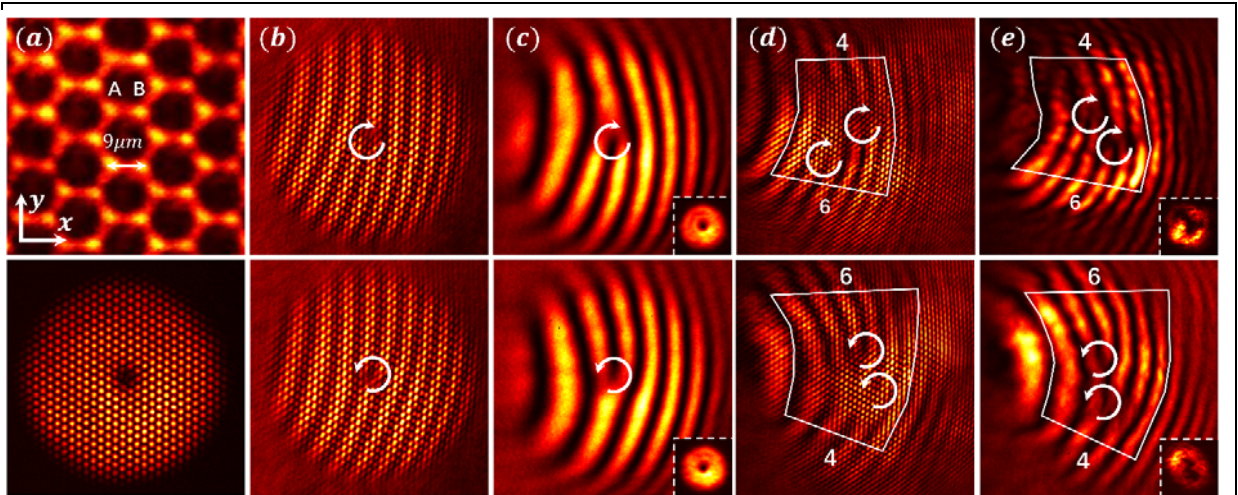

**FIG. 2: Experimental demonstration of topological charge conversion in pseudospin-1/2 HCL.** (a) Top: an optically induced HCL; Bottom: input pattern of vortex bearing triangular lattice used to excite pseudospin-orbital interaction. (b-e) Top (bottom) row corresponds to initial excitation of $s = 1/2$ ($s = -1/2$) pseudospin state with vortex beams of initial topological charge $l = 1$ ($l = -1$). Interferograms of input (b, c) and output (d, e) with a tilted

reference beam showing topological charge conversion from 1 to 2 (top) and from -1 to -2 (bottom). (b, d) Interferogram from the whole beam, and (c, e) corresponding interferogram from one of the spectral components. Difference in the numbers of counted fringes from the two sides of the marked region illustrates the net topological charges at output in (d, e). White curved arrows mark the position and helicity of the vortices. Insets in (c) and (e) show singly- and doubly-charged vortex intensity patterns obtained at input and output, respectively, from one of the $K$ valleys as illustrated in Fig. 1(a).

To substantiate the above kinematical picture summarized in Eq. (2), experimental observations (Fig. 3a) are further corroborated by numerical simulations (Figs. 3b-3d) based on the paraxial Schrodinger wave equation[13]. For all simulations, the parameters are chosen close to those from experiment with the index contrast $\delta n = 2 \times 10^{-4}$. We excite pseudospin states $s = 1/2$ (top) and $s = -1/2$ (bottom), with the same input beam of topological charge $l = 1$. In numerical simulations, the output field is decomposed into each pseudospin component. From the phase structure of each component, the difference is clear: if the $s = 1/2$ component is initially excited, the unexcited $s' = -1/2$ component is converted into an $l' = 2$ vortex (Fig. 3d, top). In contrast, if the $s = -1/2$ is initially excited, the vorticity in the unexcited $s' = 1/2$ component disappears, $l' = 0$ (Fig. 3c, bottom). The vorticity of the initially excited component always remains unchanged, in accordance with Eq. (2). Note that the output intensity patterns in the lower insets of Fig. 3 have subtle difference between the two cases of excitation: the donut shape is preserved in the top panels (when both components maintain a vortex), but deforms to have a bright central spot in bottom panels (when vortex annihilation occurs in one of the components).

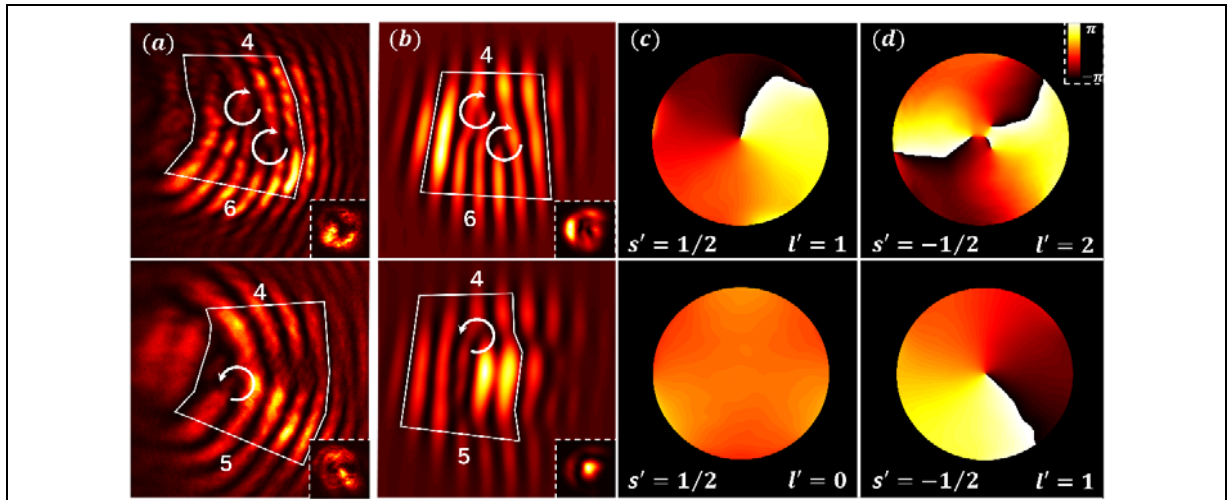

**FIG. 3: Evolution of pseudospin states in HCL.** Top (bottom) row: pseudospin state $s = 1/2$ ($s = -1/2$) is selectively excited with initial beam of topological charge $l = 1$. (a) Output interfrogram from the experiment, and (b) corresponding results from the simulation

with (c, d) showing the evolved phase structure separately for each pseudospin component. The topological charge increases (decreases) by 1 unit in the initially unexcited component as seen in the top (bottom) panel of d (c), while that in the initially excited component remains intact, as governed by Eq. (2). The lower insets in (a, b) are the corresponding intensity patterns; the doughnut (central bright spot) pattern corresponds to $l \neq 0$ ($l = 0$) components.

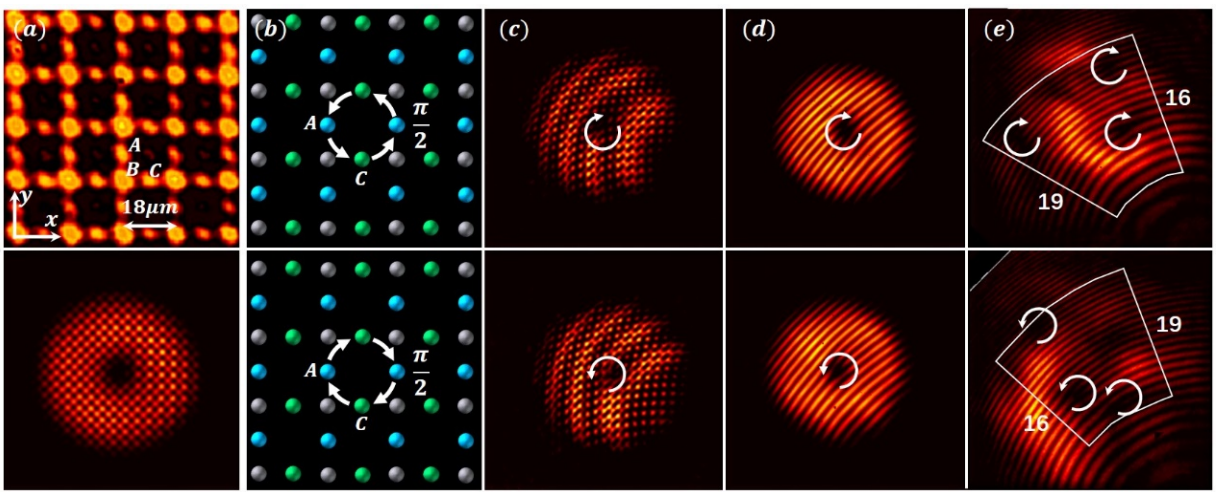

**FIG. 4: Experimental demonstration of topological charge conversion in pseudospin-1 Lieb lattice.** (a) Top: an optically induced Lieb lattice; Bottom: input pattern of vortex-bearing square lattice used to excite the Lieb lattice. (b) Top (bottom) row illustrates selective excitation of $A$ and $C$ sublattices with an appropriate phase relation optimized for pseudospin states $s = 1$ ($s = -1$) by vortex beams of initial topological charge $l = 1$ ($l = -1$). (c-e) Interferograms of input (c, d) and output (e) showing topological charge conversion from 1 to 3 (top) and from -1 to -3 (bottom).

Next, we discuss experimental results obtained with a photonic Lieb lattice (summarized in Fig. 4). The lattice is established by optical induction in a 20-mm-long crystal[36], with a nearest neighbor spacing of 9 μm. A donut-shaped square lattice beam (Fig. 4a, bottom) created by interfering four singly-charged vortex beams is employed as a probe, which excite only one pseudospin component (top panel: $l = 1, s = 1$; bottom panel: $l = -1, s = -1$). In the Lieb lattice, the pseudospin $S_z$ is not diagonal in sublattice basis[15]. Therefore, to excite a given pseudospin state, the probing square lattice is matched either to the $B$ sublattice (for the $s = 0$ pseudospin state) or the $A$ and $C$ sublattices with appropriate phase relation (for the $s = 1$ and $s = -1$ pseudospin states) as illustrated in Fig. 4b. The output interferograms in Fig. 4e clearly display three vortices of the same helicity as the input (i.e., a net topological charge of 3 or -3), exhibiting a conversion of the topological charges from $l$ to $l + 2s$ for the optimally aligned excitations (schematically illustrated in Fig. (1b).

In Fig. 5, we show experimental (Fig. 5a) and numerical (Fig. 5b) results obtained by initial excitation of the pseudospin states $s = 1$, $s = 0$, and $s = -1$ with the same input beam of topological charge $l = 1$, and examine how the phase evolves at output for each decomposed pseudospin component (Figs. 5c-5e). The first case ($s = 1$) corresponds to optimally aligned excitation in Fig. 4, where the topological charge emerged in the $s' = -1$ pseudospin component is $l' = 3$. For the latter two cases ($s = 0$, and $s = -1$), which are not optimally aligned, the initial vortex is also transformed into multiple vortices but with a net topological charge of 2 (middle) or 1 (bottom), while all components satisfy again the kinematics of pseudospin-orbit interaction [Eq. (2)]. Similar studies with the input beam of topological charge $l = -1$ led to the same conversion rule. Due to the non-diagonal nature of pseudospin-1 Hamiltonian, the pseudospin components of the Lieb lattice do not have a trivial correspondence to a particular sublattice[15] as for the case of pseudospin-1/2 HCL[13]. In fact, the physics of pseudospin-orbit interaction in Lieb lattices is in one aspect richer than that of polarization-based spin-orbit interaction: here we have excited also the $s = 0$ pseudospin state, inadmissible for helicity of photons due to its zero mass. Moreover, as expounded below, we have observed the topological conversion arising from the Berry phase of $2\pi$, which may seem trivial at glanc[37].

In what follows we develop a systematic theoretical framework to fully analyze the observed phenomena, which unravels the connections between pseudospin, OAM and the underlying topology of the lattice in $k$-space. The initial vortex beam which probes the conical intersection of Hamiltonian (1) is described with the complex amplitude of the electric field $\psi_{l,s}(r, \varphi_r, z = 0) = \psi_0 r^l e^{il\varphi_r} \exp(-r^2/a_0^2) \chi_{S,s}$, where $\chi_{S,s}$ accounts for the fact that we initially excite only a single pseudospin component. The probe beam is broad in real space ($a_0 \gg$ lattice constant), and narrow in momentum space.

We expand the initial excitation in eigenmodes of Hamiltonian (1) and account for the dynamics via $\psi_{l,s}(r, \varphi_r, z) = \sum_{n,\mathbf{k}} c_{n,\mathbf{k}} \psi_{n,\mathbf{k}} \exp(i\beta_{n,\mathbf{k}} z)$; the coefficients $c_{n,\mathbf{k}}$ are found by calculating projection $\langle \psi_{n,\mathbf{k}} | \psi_{l,s}(z = 0) \rangle$ (See SI for details). For the optimally aligned initial state in HCL (e.g., exciting the $s = 1/2$ state with $l = 1$), it is straightforward to find the evolving complex amplitude of the electric field:

$$\psi_{l=1,s=\frac{1}{2}}(r, \varphi_r, z) = e^{il\varphi_r} \chi_{\frac{1}{2},\frac{1}{2}} g_{\frac{1}{2},\frac{1}{2}}(r,z) + e^{i(l+1)\varphi_r} \chi_{\frac{1}{2},-\frac{1}{2}} g_{\frac{1}{2},-\frac{1}{2}}(r,z). \qquad (3)$$

The radial and $z$-dependence of the electric field is for clarity denoted with $g_{\frac{1}{2},\pm\frac{1}{2}}(r,z)$, and it produces the conical diffraction pattern[13, 33]. This is in full agreement with observations in Figs. 2 and 3 (top rows), which show that the initially unexcited $s' = -1/2$ component has vorticity $l' = l + 1$, whereas the excited $s' = 1/2$ component has vorticity $l' = l$.

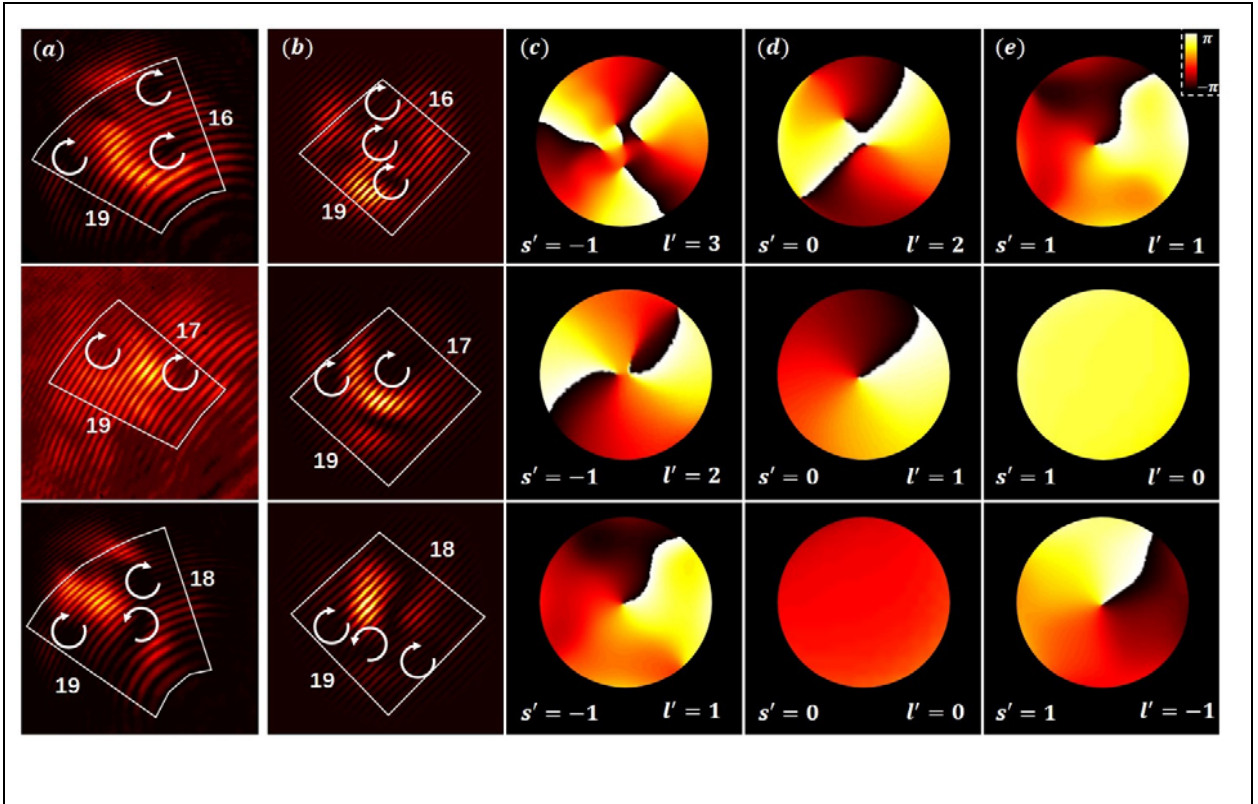

**FIG. 5: Evolution of pseudospin states in Lieb lattice.** Top to bottom rows are for initial excitations of the pseudospin state $s = 1, s = 0,$ and $s = -1,$ with same input beam of topological charge $l = 1$. Output interfrograms from (a) experiment and (b) simulation show that different topological charge conversions occur under different excitation conditions. (c-e) show the output phase structure of the probe beam decomposed for each pseudospin component $s'$, where the output vorticity $l'$ in each component has been identified. In all cases, each pseudospin component obeys Eq. (2).

In a fully equivalent manner, we can describe the dynamics in the Lieb lattices. For the optimally aligned initial excitation (e.g., exciting the $s = 1$ state with $l = 1$), the electric field evolves according to the following (see SM for details):

$$\psi_{l=1,s=1}(r,\varphi_r,z) = e^{il\varphi_r}\chi_{1,1}g_{1,1}(r,z) + e^{i(l+1)\varphi_r}\chi_{1,0}g_{1,0}(r,z) + e^{i(l+2)\varphi_r}\chi_{1,-1}g_{1,-1}(r,z). \quad (4)$$

We see that the topological charge emerged in the pseudospin components $s = 0$ and $-1$ is $l + 1$ and $l + 2$, respectively, in accordance with experimental results and numerical simulations presented in Figs. 4 and 5 (top rows), and the kinematical arguments. Dynamical considerations provide, in addition, the details of the $r$- and $z$-dependence of the electric field, which are contained in the g-functions. For all other initial conditions, similar calculations yield results also in accordance with observations.

Although the kinematical and dynamical explanations seem enough to address the observed phenomena, there is a fundamental connection to the underlying topology of $k$-space and the observation of vortices in $x$-space. If a beam propagates sufficiently long in a photonic lattice, as in our experiments, the output intensity of the beam in $x$-space will reflect the initial distribution of the power in the lattice $k$-space (analogous to far-field dynamics in free space). However, the HCL possesses topological singularity at the Dirac point: the Berry phase acquired as one traverses a loop around the Dirac point is $\pi$; the Dirac point can be considered as a flux tube (topological singularity) of the Berry curvatur[38]. Our experiments essentially reveal how the excitations of modes around the singularity are mapped into the far field dynamics.

To see that clearly, we revisit the calculation of the Berry phase in the HCL. An HCL eigenstate close to the conical intersection can be written as $\psi_{n,\mathbf{k}} = \frac{1}{\sqrt{2}}\begin{pmatrix} n \\ e^{i\varphi_k} \end{pmatrix}$, with $n = \pm 1$. As we adiabatically circle around the Dirac point, the acquired Berry phase is $i\oint \left\langle \psi_{n,\mathbf{k}} \left| \frac{\partial}{\partial \varphi_k} \right| \psi_{n,\mathbf{k}} \right\rangle = \pi$. The Berry phase arises from the specific phase relation in $k$-space between the pseudospin components of the eigenstate. There is a vortex (i.e., a topological charge) in $k$-space in one of the pseudospin components; more precisely, the difference in $k$-space topological charges of the two components is one. From the derivation of Eq. (3), we see that during propagation this vortex is mapped from the $k$-space to the $x$-space (see SM). Thus, what we observed in our experiments is the topological singularity of the HCL mapped from momentum to real space. The mapping is revealed due to the properly designed initial excitation of a single pseudospin state.

This finding holds for a general Hamiltonian (1), i. e, for any pseudospin $S$. Every eigenmode can be expanded in pseudospin eigenstates as $\psi_{n,\mathbf{k}} = \sum_{s=-S}^{S} \langle \chi_{S,s} | \psi_{n,\mathbf{k}} \rangle \chi_{S,s}$. The coefficients $\langle \chi_{S,s} | \psi_{n,\mathbf{k}} \rangle$ are found by rewriting the Hamiltonian as $H = \frac{\kappa}{2} k \left( S_+ e^{-i\varphi_k} + S_- e^{i\varphi_k} \right)$, where $S_\pm = S_x \pm i S_y$:

$\beta_{n,\mathbf{k}} \langle \chi_{S,s} | \psi_{n,\mathbf{k}} \rangle = \langle \chi_{S,s} | H | \psi_{n,\mathbf{k}} \rangle =$
$\frac{\kappa}{2} k \left( \sqrt{(S-s)(S+s+1)} \langle \chi_{S,s+1} | \psi_{n,\mathbf{k}} \rangle e^{-i\varphi_k} + \sqrt{(S+s)(S-s+1)} \langle \chi_{S,s-1} | \psi_{n,\mathbf{k}} \rangle e^{i\varphi_k} \right).$
(5)

There is a clear phase relationship between different pseudospin components of the eigenstates. The difference in $k$-space topological charges (vortices) of neighboring pseudospin components is one. When a single pseudospin component is excited, $k$-space topological charges of the unexcited components are mapped to real space, which is the fundamental mechanism behind topological charge conversions illustrated in Figs. 1-5.

A few issues merit further discussion. The first is about the theoretical explanation of experimental results by utilizing conservation of angular momentum, $[J_z, H] = 0$. Surprisingly, the topological charge conversion from $l$ to $l + 2s$ holds also in stretched lattices, where conical intersections are described by the Hamiltonian $H_s = \kappa_x S_x k_x + \kappa_y S_y k_y$, and the angular momentum is not conserved, $[J_z, H_s] \neq 0$, due to lack of rotational symmetry. An inspection of the eigenstates of $H_s$ for pseudospin $S = 1/2$ and $S = 1$ shows that the $k$-space vortices become elliptical but preserve their topological charge. For the stretched HCL, the winding of the Berry phase around the Dirac point is topologically protected, until the stretching is sufficiently large so that the inequivalent Dirac points merge and a gap opens[39-41]. The topological charge

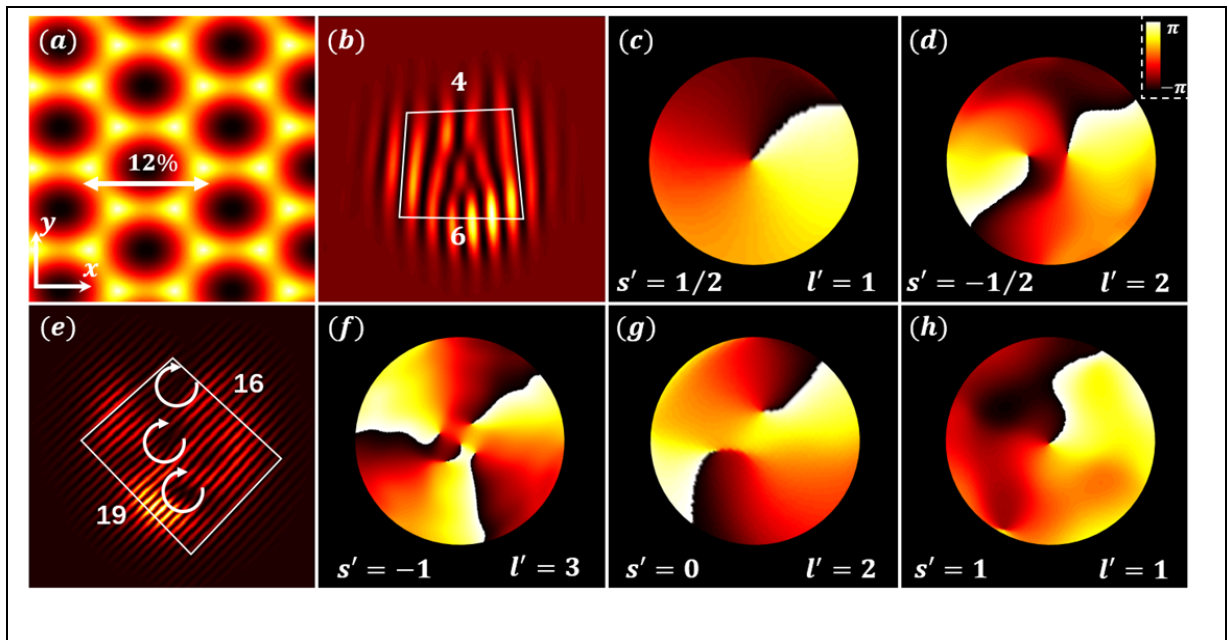

**FIG. 6: Evolution of pseudospin states in stretched lattices lacking rotational symmetry around conical intersections.** Top (bottom) row corresponds to a stretched HCL (Lieb lattice), with an initial excitation $l = 1, s = 1/2$ ($l = 1, s = 1$). (a) A 12% horizontally stretched HCL. (b, e) Interferograms of the output beam, which clearly indicate the topological conversion from $l$ to $l + 2s$, corresponding to results from unstretched lattices of the top rows in Fig. 3b and Fig. 5b. This is underpinned by the phase structure of the pseudospin

components at the output illustrated in (c,d) and (f,g,h) for the two lattices.

conversion no longer occurs after this transition. In Fig. 6 we show numerical simulations for the optimally aligned initial condition in the stretched HCL and Lieb lattice. The nearest neighbor spacing for the HCL and the Lieb lattice is 9 μm, stretched by 12% and 15%, respectively. As seen from these results, the conversion of the topological charges from $l$ to $l + 2s$ holds even when the angular momentum $J_z$ is not conserved, which indicates that the mapping of the topological charges from $k$-space to $x$-space is a fundamental process with topological origin.

The second is about going beyond the 2D platform. Our experiments are performed in 2D lattices, however, analogous considerations can be made for 3D Hamiltonians. For example, we consider the Weyl Hamiltonian $H_{Weyl} = \boldsymbol{\sigma} \cdot \mathbf{k}$, which has attracted considerable interest in recent years; it has been experimentally realized in the Brillouin zone of specially designed optical and condensed matter structures[16,17]. This Hamiltonian is a synthetic magnetic monopole in momentum space [18], with topological charge 1. Suppose that we initially excite the modes around a Weyl point with a symmetric Gaussian-like distribution, and that we excite a pseudospin eigenstate $\chi_{\frac{1}{2},\frac{1}{2}}$. The initial state is $\psi_{0,\frac{1}{2}}(r, \theta_r, \varphi_r, t = 0) = \psi_0(r)\chi_{\frac{1}{2},\frac{1}{2}}$, and its evolution in time is governed by $i\frac{\partial \psi}{\partial t} = H_{Weyl}\psi$. By expanding the initial state in eigenmodes of the Weyl Hamiltonian, it is straightforward to see that the wavefunction evolves as (see SM)

$$\psi_{0,\frac{1}{2}}(r, \theta_r, \varphi_r, t) = \chi_{\frac{1}{2},\frac{1}{2}} g_{\frac{1}{2},\frac{1}{2}}(r, \theta_r, t) + e^{i\varphi_r} \chi_{\frac{1}{2},-\frac{1}{2}} g_{\frac{1}{2},-\frac{1}{2}}(r, \theta_r, t). \tag{6}$$

Clearly, even though the initial state was a Gaussian-like excitation with $l = 0$ and $s = 1/2$, in the unexcited pseudospin component $s' = -1/2$ a vortex with topological charge $l' = 1$ emerges, as illustrated in Fig. 1c. Since this is a 3D rotationally invariant Hamiltonian, $[\mathbf{J}, H_{Weyl}] = 0$, if we excite any pseudospin state $\alpha\chi_{\frac{1}{2},\frac{1}{2}} + \beta\chi_{\frac{1}{2},-\frac{1}{2}}$ with a Gaussian-like distribution, we will obtain in the output a vortex field with a topological charge identical to the charge of the Weyl monopole, with vorticity pointing in the direction of the initial pseudospin. Thus, with properly designed initial excitation, mapping of topological properties of the Weyl monopole to topological charges in real space can be readily achieved. In fact, this type of dynamics in 3D Weyl systems, achievable in ultracold atomic gases[19], seems to be related to a recent experiment where an electron beam scattered from a magnetic monopole experienced a conversion into an electron vortex[20].

We have thus demonstrated the universal mapping of topological singularities (topological charges) in momentum space to topological entities in real space. The mapping is most clearly revealed for an optimally aligned excitation of pseudospin states that leads to topological charge conversion obeying the rule $l \rightarrow l + 2s$. As a fundamental phenomenon, such a mapping retains for all other excitation conditions and for any arbitrary spin $s$, and upholds even in a stretched lattice system where the angular momentum is not necessarily conserved. We have predicted that, driven by the topological origin, the same mechanism exists in 3D Weyl lattices where synthetic magnetic monopoles come to play the role. Our findings bring about many interesting questions as well as opportunities. For instance, is it possible to create vortices of Bose-Einstein condensates[42] by topological conversion from synthetic magnetic monopoles in ultracold atomic gases[19]? How could the mechanism explored here be adapted for topological conversion with photons in a photonic Dirac monopole field[43]? It is also natural to ask: is the spin angular momentum gifted by light polarization indispensable in spin-to-orbital angular momentum conversion, as commonly thought, or the pseudospin and topological conversion is essential even in those conventional settings based on optical phase elements [24-26]? What other mechanisms can we conceive and explore where topological properties of the bands can be directly mapped from momentum to real space in experiments? Can other topological entities such as vortex knots and nodal chains[44-46] be directly mapped from momentum space to real space or vice versa, or onto a synthetic space[47, 48]?

## Acknowledgements

This research is supported by the National key R&D Program of China under Grant (No. 2017YFA0303800), the National Natural Science Foundation (91750204, 11674180), PCSIRT, and the 111 Project (No. B07013) in China. H.B. acknowledges support in part by the Croatian Science Foundation Grant No. IP-2016-06-5885 SynthMagIA, and the QuantiXLie Center of Excellence, a project co-financed by the Croatian Government and European Union through the European Regional Development Fund - the Competitiveness and Cohesion Operational Programme (Grant KK.01.1.1.01.0004). D.L. is supported by the Institute for Basic Science in Korea (IBS-R024-Y1).